\documentclass[twoside]{ima} 
\usepackage{amsmath,amssymb}

\def\cocoa{{\hbox{\rm C\kern-.13em o\kern-.07em C\kern-.13em o\kern-.15em A}}}

\newcommand{\indep}{ \_\hskip-4pt \sqcup \hskip-5pt\_}

\begin{document}
\markboth{EVA RICCOMAGNO AND JIM Q. SMITH}{ALGEBRAIC CAUSALITY: BAYES NETS AND BEYOND}        

\title{ALGEBRAIC CAUSALITY: BAYES NETS AND BEYOND}

\author{EVA RICCOMAGNO\thanks{
    	Department of Mathematics, Universit\`a degli studi di Genova, Via Dodecaneso 35, 16146, Italy
    	(riccomagno@dima.unige.it). }
  \and JIM Q. Smith\thanks{Department of Statistics, The University of Warwick, Coventry, CV4 7AL, UK
        (j.q.smith@warwick.ac.uk).}}

\maketitle   
             
\begin{abstract}                   
The relationship between algebraic geometry and the inferential framework
of the Bayesian Networks with hidden variables has now been fruitfully explored and exploited
by a number of authors. More recently the algebraic formulation
of Causal Bayesian Networks has also been investigated in this context. After reviewing
these newer relationships, we proceed to demonstrate that many
of the ideas embodied in the concept of a ``causal model'' can be 
more generally expressed directly in terms of a partial order and a family of
polynomial maps. The more conventional
graphical constructions, when available, remain a powerful tool.  
\end{abstract} 
 
\begin{keywords} 
Bayesian networks, causality, computational commutative algebra.
\end{keywords}

{\AMSMOS 
\endAMSMOS}

\section{Introduction}
		
		There has been much recent interest in the study of causality based on
		graphs, e.g. \cite{Dawid02,Pearl2000,Pearl2003,Spirtesetal93}. 
		A most common scenario studied is when the observer
		collects data from a system and wants to make inferences about what would
		happen were she to control the system, for example by imposing a new treatment
		regime. To make prediction with such data she needs to hypothesize 
		a certain causal mechanism which not only describes the data generating process, 
		but also governs what might happen were she to control the system. 
		Pioneering work by two different groups of authors \cite{Pearl2000, Spirtesetal93} 
		have used a graphical framework called a Causal Bayesian
		Network (CBN). Their work is based on Bayesian Networks (BN) which is a compact framework for representing 
		certain collections of conditional independence statements. 

Algebraic geometry and computational commutative algebra have been successfully 
employed to address identifiability issues \cite{GarciaStillmanSturmfels2005,Mond,SettimiSmith2000}
and to understand the properties of the learning mechanisms \cite{Riccimpu,RiccoSmith,Riccsmith07}
behind BN's. 
A key point was the understanding that collections of conditional independence relations on discrete random variables
expressed in a suitable parametrization are polynomials and have a close link with toric varieties \cite{GarciaStillmanSturmfels2005,Mono}.
Further related work showed that pairwise independence and global independence are expressed through toric ideals \cite{GeigerMeekSturmfels2006} and that Gaussian BN's are related to classical constructions in algebraic geometry e.g. \cite{Sullivant2007}.

In this paper we observe that when 
model representations and causal hypotheses are expressed as a set of maps from one semi-algebraic space to
another, then ideas of causality are separated from the classes of graphical models. 
This allows us to generalise straightforwardly concepts of graphical causality as defined in e.g. \cite[Definition 3.2.1]{Pearl2000} to non-graphical model classes.  
Many classes of models including context specific BN's \cite{Freidman,McAllister,Poole,Salmaron}, 
Bayes Linear Constraint models (BLC's) \cite{Riccimpu} and Chain Event Graphs (CEG's) 
\cite{Smith07,Riccsmith07,ThwaitesPrague,ThwiatesParis}  
are special cases of this algebraic formulation.

Causal hypotheses are most naturally expressed in terms of two types of hypotheses. 
The first type concerns when and how circumstances might unfold. This provides 
us with a hypothesized partial order which can be reflected by the 
parametrization of the joint probability mass function of the idle system. 
The second type of hypotheses concerns structural assertions about the uncontrolled system that,
we assume, also apply in the controlled system. These are usually expressible as semi-algebraic constraints in the given parametrization.
Under these two types of hypotheses the mass function of the manipulated system is defined as a projection 
of the mass function of the uncontrolled system, in total analogy to CBN's. 
The combination of the partial order and of these constraint equations and inequalities 
enables the use of various useful algebraic methodologies for the 
investigation of the properties of large classes of  discrete inferential models 
and of their causal extensions. 

The main observation of the paper is that a (discrete) causal model can be redefined directly and very flexibly using an algebraic representation
starting from a finite set of unfolding events and a description of a way they succeed one another. 
This is shown through model classes of increasing generality. 
First in Section \ref{Section2} we review the popular class of discrete BN models, our simplest class,
their related factorization formulae under a preferred parametrisation, and
their causal extensions. Then we extrapolate the algebraic features of BN and
give their formalisation in   
Section~\ref{Section3} in a rather general context. 
In Section \ref{Section4} we show how this formalization can apply to more general
classes of models than BN's, so that identifiability and feasibility issues can be
addressed. Here we describe causal models based on trees in Section \ref{SectionTree} 
and the most general model class we consider is in Section \ref{extreme}.

The issues are illustrated throughout by a typical albeit 
simple model for the study of the causal effects of violence of men who might watch a violent
movie, introduced in Section \ref{Section2violent} 
to outline some limitations of the framework of the BN for
examining causal hypotheses, which, we believe, currently is the best framework to represent causal hypotheses.
In Section \ref{Section4.1} we are able to express these limitations within an algebraic setting. 

\section{Notes on causal Bayesian networks} \label{Section2}

\subsection{The BN and its natural parametrization} \label{Section2BN}

The discrete BN is a powerful framework to describe hypotheses an observer might make about a particular system. 
It consists of a directed acyclic graph with $n$ nodes and of a set of probabilistic statements. 
It implicitly assumes that the features of main interest in a statistical model can be expressed in the
following terms. 

\begin{itemize}
\item The observer's beliefs as expressed through the graph concern statements
about relationships between a prescribed set of measurements $\boldsymbol{X}=\left\{
X_{1},X_{2},\ldots X_{n}\right\} $ taking values $\left\{ x_{1},x_{2},\ldots
x_{n}\right\} $ in a product space $S_{X}=\mathbb{X}_{1}\times \mathbb{X}
_{2}\times \ldots \times \mathbb{X}_{n}$, where $X_i$ is a random variable that takes values in $\mathbb{X}_i$, $1\leq i\leq n$. 
For $1\leq i\leq n$ let $r_{i}$ be the cardinality of $\mathbb{X}_{i}$, $r_i$ be finite 
and $X_{i}$ take value on the set of integers $1,2,\ldots ,r_{i}$, henceforth indicated as $[r_i]$.
Then the joint sample space 
$S_{X}$ contains $r=\prod_{i=1}^{n}r_{i}$ distinct points.

\item The sets of relationships most easily read out of the graph, are consistent
with a partial order $\prec$ on $X_{1},X_{2},\ldots X_{n}$ implied by the graph itself.
Historically this order was often chosen so that if $1 \leq i_1 < i_2 \leq n$ then 
$X_{i_1} \prec X_{i_2}$ in some rather loose mechanistic sense,
although this is certainly not a necessary interpretation of the order. In this case
we will call the BN \emph{regular}. Henceforth we will assume a regular BN.

\item The graph expresses the $n-1$ conditional independence statements 
\[
X_i \indep \{X_1, X_2, \ldots, X_{i-1} \} \setminus Pa(X_i) | Pa(X_i)
\] 
where  $Pa(X_i)$ is called the \emph{parents} of $X_i$. For a definition see \cite{lauritzen1996}.
For $1 \leq  i \leq n$, in some sense the values the random variables in $Pa(X_i)$ take, embody all relevant
probabilistic information concerning $X_i$. Furthermore 
for regular BN's $Pa(X_i)$ can be interpreted as the set of variables in $\boldsymbol{X}$ relevant to the potential development of
$X_i$.
\end{itemize}

The last property enables the entire set of beliefs to be expressed by a single
directed acyclic graph called a BN. 
Its vertex set is the set of measurement variables $\{ X_1,X_2, \ldots, X_n\}$ 
and there is an edges from $X_j$ to $X_i$ if and only if $X_j \in Pa(X_i)$. The implicit
partial order induced by this direct graph and its loose link to the order of how
circumstances unfold, has encouraged various authors to extend the model to
one that also makes statements about relationships between the same set of
measurements when they have been subjected to various controls, e.g. \cite{Pearl2000,Spirtesetal93}. 
Before discussing this point, we consider an example  to underline some specific features. 

\subsubsection{A violent example} \label{Section2violent}

Consider a statistical model built to study whether watching a violent movie might
induce a man into a fight, allowing for testosterone levels to, at least partially,
explain a violent behaviour. 
Let $X_{2}$ denote whether a man watches a violent movie early one evening 
$\{x_{2}=1\}$ or not $\{x_{2}=2\}$ and let $X_{4}$ be an 
indicator of whether he is arrested for fighting $\{x_{4}=1\}$ or not $
\{x_{4}=2\}$ late that evening. 
If he watches the movie, let $X_{1}$ denote his testosterone level just before seeing it and 
$X_{3}$ his testosterone level late that evening. For a man who does not watch the movie let 
$X_{1}=X_{3}$ denote his testosterone level that evening. 

Assume $X_1$ and $X_3$ take three values: $1$ for low levels of testosterone, 
$2$ for medium levels and $3$ for high levels, so that $
\left( r_{1},r_{2},r_{3},r_{4}\right) =\left(
3,2,3,2\right)$ and $r=36$. Then this can be depicted as the following BN
\[
\begin{array}{ccc}
X_{1} & \rightarrow & X_{3} \\ 
& \nearrow & \downarrow \\ 
X_{2} & \rightarrow & X_{4}
\end{array}
\]
The graph of this BN embodies two substantive statements. The first one, 
$X_{2}\indep X_{1}$ is associated with the missing edge from $X_{1}$ to 
$X_{2}$ and states that whether 
the man watched the movie would not depend on his testosterone level.
The second one $X_{4}\indep X_{1}| (X_{2},X_{3})$  is associated with the missing edge 
from $X_{1}$ to $X_{4}$ and states that
the testosterone level before watching the movie gives no additional relevant
information about the man's inclination to violence provided that we
happen to know both whether he watched the movie and his current
testosterone levels. It will be useful later to note that the edge $(X_2,X_3)$ indicates that watching a violent movie might
help cause the fight by increasing testosterone levels, 
while the edge $(X_2,X_4)$ indicates that it might do so by some other mechanism.

An alternative semi-algebraic representation of this statistical model is given as follows. 
For each of the $r=36$ levels $\boldsymbol{x}=(x_1,x_2,x_3,x_4)\in S_{\boldsymbol{X}}$ let 
$p(\boldsymbol{x})=\operatorname{Prob}(X_1=x_1,\ldots,X_4=x_4)$ be the joint mass function associated with the BN.
For the sake of simplicity we assume $p(\boldsymbol{x})$ strictly positive for each $\boldsymbol{x}$. 
An obvious inequality constraint is given by the fact that the vector $(p(\boldsymbol{x}): \boldsymbol{x}\in S_{\boldsymbol{X}})$ lies in the 
standard simplex  
\begin{equation}
\begin{array}{rcl}
\Delta _{r-1} & = & \{u\in \mathbb{R}^{r}: 
\sum_{i=1}^{r}u_{i}=1 \text{ and } u_{i}\geq 0\text{ for }i=1,\ldots ,r\}  .  
\end{array}
\label{simplexnotation}
\end{equation}

The BN suggests the partial order on the variables for which $X_1$ and $X_2$ precede $X_3$ which precedes $X_4$. 
A natural, not unique, parametrization is, then, determined by the total ordered sequence $ X_{1}, X_{2},X_{3}, X_4$ and has 63 parameters:
$\pi_{1}(x_{1})=\operatorname{Prob}(X_1=x_1)$, $\pi_{2}(x_{2}|x_{1})= \operatorname{Prob}(X_2=x_2| X_1=x_1)$,
$\pi_{3}(x_{3}|x_{1},x_{2})=\operatorname{Prob}(X_3=x_3|X_1=x_{1},X_2=x_{2})$ and
$\pi _{4}(x_{4}|x_{1},x_{2},x_{3})=\operatorname{Prob}(X_4=x_4|X_1=x_{1},X_2=x_{2}, X_3=x_3)$.
Call the indeterminates  $\pi _{1}(x_{1}),\pi _{2}(x_{2}|x_{1}),\pi
_{3}(x_{3}|x_{1},x_{2}),\pi _{4}(x_{4}|x_{1},x_{2},x_{3})$ 
 \emph{primitive probabilities}, for $(x_1,x_2,x_3,x_4)\in S_{\boldsymbol{X}}$.
Sum-to-one constraint like $\sum_{x_2=1,2}\pi_{2}(x_{2}|x_{1})=1$ gives 28 linear constraints 
to be coupled with the positivity assumption. 

A general joint mass function on $(X_1,X_2,X_3,X_4)$ is given by the 36 quartic equations
\begin{equation}
p(\boldsymbol{x})=\pi _{1}(x_{1})\pi _{2}(x_{2}|x_{1})\pi
_{3}(x_{3}|x_{1},x_{2})\pi _{4}(x_{4}|x_{1},x_{2},x_{3})  .  \label{BNfactEs}
\end{equation}
This is a particular form of the general factorisation of the joint mass function with respect to a BN
\begin{equation}
\operatorname{Prob}(X=\boldsymbol{x})=\prod_{i=1}^n \pi(x_i|Pa(X_i)=pa(x_i))
\label{BNfact}
\end{equation}
where $\boldsymbol{x}\in S_{\boldsymbol{X}}$, $\pi(x_i|Pa(X_i)=pa(x_i))=\operatorname{Prob}(X=x_i|Pa(X_i)=pa(x_i))$
 and $pa(x_i)$ is the value taken by the random vector $Pa(X_i)$ when $X=\boldsymbol{x}$.

The conditional independence statements in the BN are given by a finite set of
linear equations in primitive probabilities
\begin{eqnarray}
\pi _{2}(x_{2}|x_{1}) &=&\pi _{2}(x_{2}|x_{1}^{\prime })\triangleq \pi
_{2}(x_{2})\text{ (say)}  \label{BNeqs} \\
\pi _{4}(x_{4}|x_{1},x_{2},x_{3}) &=&\pi _{4}(x_{4}|x_{1}^{\prime
},x_{2},x_{3})\triangleq \pi _{4}(x_{4}|x_{2},x_{3})\text{ (say)}  \nonumber
\end{eqnarray}
for all $x_{1},x_{1}^{\prime }=1,2,3$. See \cite{DawidStudeny1999} for a proof and a discussion of this. 
The statistical model expressed by the BN is then given as a semi-algebraic set 
defined by polynomial equations and inequalities in the primitive probabilities.

Furthermore the simple substitution of Equations (\ref{BNeqs}) into (\ref{BNfactEs}) allows us 
to reduces the number of parameters and of constraints. 
Indeed the resulting vectors  $(\pi _{1}(1),\pi _{1}(2),\pi _{1}(3))$
lie in $\Delta _{2}$ as do each of the vectors $(\pi
_{3}(1|x_{1},x_{2}),\pi _{3}(2|x_{1},x_{2}),\pi _{3}(3|x_{1},x_{2}))$ for $
x_{1}=1,2,3$ and $x_{2}=1,2$ whilst the vectors $(\pi _{2}(1),\pi _{2}(2))$
and each of the vectors $(\pi _{4}(1|x_{2},x_{3}),\pi _{4}(2|x_{2},x_{3}))$
for $x_{2}=1,2$ and $x_{3}=1,2,3$ lies in $\Delta _{1}$. 
Each of the 14 simplices also embodies a linear
constraint through its sum-to-one condition making the interior of the
domain a 21 dimensional linear manifold.

A critical point to notice for the generalisations that follow is that each
of the $14$ simplices $(\pi _{i}(x_{i}|Pa(X_{i})):x_{i}\in \mathbb{X}_{i})$
is labelled by a particular configuration of $Pa(X_{i})$.
In a BN each such configuration of parents labels and distinguishes a possible
history of circumstances and might influence the probabilistic
development of the network. 

It is common for a statistical model to contain as its substantive
hypotheses more than the conditional independence statements,
expressible in a BN. Often such additional
non-graphical hypotheses can be expressed as a set of 
algebraic equations or inequalities on the primitive probabilities. 
We list a few such additional hypothesis for our example.
\begin{itemize} 
\item  If the movie is not watched then we would expect 
$X_{3}=X_{1}|(X_{2}=2)$, equivalently
\begin{equation}
\pi _{3}(x_{3}|x_{1},x_{2}=2)=\left\{  
\begin{array}{ll} 1 & \text{ if } x_{3}=x_{1} \\
0 & \text{ otherwise.}  \end{array} 
\right. \label{nomovie}
\end{equation}
\item If a unit did watch the movie, we would not expect this to reduce his testosterone level.
This sets some of the primitive probabilities to zero, namely
\begin{equation}
\begin{array}{c|ccc}
X_{3}|X_{1}=x_{1},X_{2}=1 & x_{3}=1 & x_{3}=2 & x_{3}=3 \\ 
\hline
x_{1}=1 & \pi _{3}(1|1,1) & \pi _{3}(2|1,1) & \pi _{3}(3|1,1) \\ 
x_{1}=2 & 0 & \pi _{3}(2|2,1) & \pi _{3}(3|2,1) \\ 
x_{1}=3 & 0 & 0 & 1
\end{array}
\label{movietable}
\end{equation}
\item The assumption that the higher the prior 
testosterone levels the higher the posterior ones, is given by  
\begin{eqnarray}
\pi _{3}(1|2,1) &=&r_{3,2}\pi _{3}(1|1,1)  \label{movieineq} \\
\pi _{3}(3|2,1) &=&r_{3,3}\pi _{3}(3|1,1)  \nonumber
\end{eqnarray}
where $0\leq r_{3,2},r_{3,3}\leq 1$ are 
additional semi parametric parameters.  
\item
Similarly it is reasonable to expect that higher levels of testosterone
together with having seen the movie would make more probable that a man would be
arrested for fighting. This can be expressed as
$
\pi _{4}(1|1,x_{3})=r_{4,x_{3}}\pi _{4}(1|2,x_{3})
$ for $x_{3}=1,2,3$ and for $x_{3}=1,2$ 
$
\pi _{4}(1|1,x_{3}+1) =r_{4,x_{3}}^{\prime }\pi _{3}(1|1,x_{3}) 
$ and
$\pi _{4}(1|2,x_{3}+1) =r_{4,x_{3}}^{\prime \prime }\pi _{3}(1|2,x_{3})
$
where $0\leq r_{4,x_{3}},r_{4,x_{3}}^{\prime },r_{4,x_{3}}^{\prime \prime
}\leq 1$, similarly to the previous bullet point.
\item Finally a common simple log-linear response model might assume 
$r_{4,1}=r_{4,2}=r_{4,3}$.
\end{itemize}
The point here is not that these supplementary equations and
inequalities provide the most compelling model, but rather that
embellishments of this type, whilst not graphical, are common, 
are easily expressed in the primitive probability parametrization, 
and often have an almost identical type of algebraic description
as the BN. 

In general then, a BN is a collection of monomials in
primitive probabilities and the $p(\boldsymbol{x})$ parameters. It is
defined through a total order of variables ---in the example Equations (\ref{BNfactEs})--- 
supplemented by the set of linear equations on the primitive probabilities
\[
\pi _{i}(x_{i}|x_{1},x_{2},\ldots x_{i-1})=\pi _{i}(x_{i}|x_{1}^{\prime
},x_{2}^{\prime },\ldots x_{i-1}^{\prime })
\]
whenever $(x_{1},x_{2},\ldots x_{i-1})$ and $(x_{1}^{\prime },x_{2}^{\prime
},\ldots x_{i-1}^{\prime })$ take the same value on $Pa(X_{i}),$ $1\leq
i\leq n$. In the example these are Equations (\ref{BNeqs}). More detailed types of model
specification are given by the saturated
model, e.g. Equations (\ref{BNfactEs}), supplemented by further algebraic and semi-algebraic equations 
analogous to Equations (\ref{BNeqs}) and to those in the bullet points above. 
So a strong case can be made for \emph{starting} with this class of algebraic description and
relegating the graphical formulation as a useful depiction of a particular
subclass of these structures.

The BN has other associated factorization formulae based on its clique structure, see
e.g. \cite{Cowellbook}, that are more symmetric and have been used as a
vehicle for a different algebraic formulation, see e.g. \cite{GarciaStillmanSturmfels2005,GeigerMeekSturmfels2006}. 
In fact it is often elegant to express this discrete model in terms of its natural exponential parametrization \cite{DrtonSullivant2007}. 
However for causal models the partial order on the $X_i$'s given by the topology of the BN 
---and hence the associated factorization of the joint mass function--- is critical to the definition of
the predicted effect of manipulating the system: see below. In causal modelling we have therefore found it to be more expedient
to parametrize a model directly through conditional probabilities chosen so
they are consistent with such a causal partial order.
Under the parametrization given by these primitive probabilities, a BN can be thought of as a labelling of a
collection of simplices about what might happen (the
value a node random variable might take), given the relevant past (the particular configuration of values taken by its parents).

\subsection{Manifest and hidden variables}\label{SectionManifest}

Typically it is required to infer the value of a vector $\boldsymbol{f}(p(\boldsymbol{x}):\boldsymbol{x}\in \mathbb{X})$. If we are interested in the whole joint mass function, $\boldsymbol{f}$ is the identity. Often $\boldsymbol{f}$ is a polynomial or a rational polynomial function in the primitive probabilities. Obviously such inference would be
trivial if we could learn the full probability table $p(\boldsymbol{x}):\boldsymbol{x}\in \mathbb{X}$. However usually only variables in a subset $M$ of $\{X_{1},X_{2},\ldots ,X_{n}\}$ are measured in a particular population, sometimes over a very large sample of individuals. The random variables in $M$ are called \emph{manifest} and those in $H=\{X_{1},X_{2},\ldots ,X_{n}\}\backslash M$ are called \emph{hidden}. Almost always we can learn only the values of the polynomials 
\begin{equation}
\sum_{\boldsymbol{x}_{i}\in S_H}p(\boldsymbol{x})=q(\boldsymbol{x}(M))
\label{man1}
\end{equation} 
where $\boldsymbol{x}(M)$ is a sub-vector of $\boldsymbol{x}$ involving only values of the manifest random
variables. For the example in Section \ref{Section2violent}
it may be impossible to determine the testosterone levels $H=\{
X_{1}\}$ of the individuals in any sample, but only $M=\{X_{2},X_{3},X_{4}\}$. 
If we ignore the positivity conditions, this is a
Newtonian problem in albeit real algebraic geometry and so solvable through
techniques like elimination theory. Indeed when $\boldsymbol{f}$ is the
identity these identifiability questions are now answered for many small
BN's by using elimination techniques. See e.g. \cite{GarciaStillmanSturmfels2005} and \cite{PachterSturmfels2005} 
for examples from the field of computational biology. 

Often the study of identifiability issues after observing the manifest margins (\ref{man1}) has been driven more
by the semantics of the graph of a BN where a full node of the graph represents a hidden \emph{variable}/measurement. However in practice  
missingness of data is often contingent on what has happened to a unit, i.e.
the particular \emph{value} its parent configuration takes and not the whole variable.

To illustrate this point consider collecting data for the example in Section \ref{Section2violent} when $X_4$ is hidden and it is the variable of central interest with its associated probabilities $\pi_4(x_4|x_1,x_2,x_3)$. 
It might be possible to randomly sample men and measure their testosterone levels before and after watching a violent movie.
Call this Experiment 1.
However if it were seriously believed that watching a violent movie might induce a fight, it would be
unethical to release the subjects after watching the movie, while any therapy either in the form of drugs or counselling will corrupt the experiment.
In any case recording the
proportions of subjects who later fought would not give an
appropriate estimate of probabilities associated with $X_{4}$ and
conditional on its parents. So values like $\pi_4(2|x_1,1,x_3)$ cannot be estimated from such samples. 
To identify the system we therefore need to
supplement this type of experiment with another measuring willingness to
fight. Other experiments might be envisaged leading to analogues problems.

Partial information about the joint distribution of $X_{4}$ with other variables might 
be obtained from a random sample of men arrested for fighting $\{x_{4}=1\}$. Their 
current testosterone levels $X_{3}$ and whether they had recently
watched a violent movie $X_{2}$ could be measured. But we could not measure $(X_2,X_3)$
for men that are not caught fighting. 
Thus the finest partition of probabilities we could hope for in a population
under this kind of survey is based on the sample space partition
$\{\overline{A},A(x_{2},x_{3} \}:x_{2}=1,2,x_{3}=1,2,3\}$ where 
$
\overline{A} =\{\boldsymbol{x}:X_{4}=2\}$ and 
$A(x_{2},x_{3}) = \{\boldsymbol{x}:X_{2}=x_{2},X_{3}=x_{3},X_{4}=1\}
$ 
i.e. 
$
q(\overline{A})=\sum_{\boldsymbol{x}_{i}\in \overline{A}}p(\boldsymbol{x})
$ 
and for $x_{2}=1,2$ and $x_{3}=1,2,3,$ 
$
q(A(x_{2},x_{3}))=\sum_{\boldsymbol{x}_{i}\in A(x_{2},x_{3})}p(\boldsymbol{x})$. Call this Experiment~2.

The algebraic expression of the observations from this second experiment are 
analogous to Equations (\ref{man1}), being sums of the probabilities on the atoms of the
joint mass function, but they are not of the same form because manifest 
equations do not correspond to marginal constraints. Nevertheless the types
of elimination techniques applicable to BN can clearly still be employed 
to determine the geometry and properties of its solution spaces.
So the pattern of missing data encountered often have an algebraic but not a graphical representation.

\subsection{Causal functions} \label{SectionCausal}

As already mentioned, the regular BN in Section \ref{Section2violent} could be hypothesised to be 
causal following many authors e.g. \cite{Pearl2000,Spirtesetal93}. Here the term 
``cause'' has a very specific meaning and the causal structure is conventionally associated to the partial order of the graph
in a regular BN. A formal definition is given in the next section. See also \cite[Equation (3.10)]{Pearl2000}. First we discuss some key points.

Asserting that the BN in Section \ref{Section2violent} is a CBN implies that since $X_{1}\prec X_{3}$ and $X_{2}\prec X_{3}$ we believe 
$X_{1}$ and $X_{2}$ are potential causes of $X_{3}$.
This means that if the prior level of testosterone $X_{1}$ were to be controlled to take the value 
$x_{1}$ and the man were \emph{made} to watch the film (or not to), then the
probability he had a testosterone value $X_{3}=x_{3}$ would be the
same as the proportion of times $X_{3}=x_{3}$ was observed to occur in the uncontrolled
(infinite) population with observed values $X_{1}=x_1$ and $X_2=1$ ($X_2=2$ if he was forced not to watch the movie).

Similarly, a causal interpretation of this BN would also assert that the effect on
the probability the man would fight $\{X_{4}=1\}$ if we forced $\{X_{i}=x_{i}:i=1,2,3\}$ would be 
identified with $\pi_{4}(1|x_{1},x_{2},x_{3})=\pi _{4}(1|x_{2},x_{3})$, i.e. the corresponding
conditional probability in the uncontrolled system.

Furthermore, forcing a variable to take a value $X_i=x_i$ will have no effect on the joint distribution of the variables which do not follow $X_i$ in the causal partial order. For example increasing the testosterone level $X_{3}$ would have no effect on the joint probability of $(X_1,X_2)$. 

Obviously a CBN makes stronger statements than a BN with the same graph.
As in the example above the extra modelling statements made in a CBN
are often plausible and gives us a framework within which to make
predictions about the observed system were it to be subject to certain
controls. For example we might want to consider the potential
effect of
\begin{enumerate}
\item banning the film, thus preventing it from being viewed by the general public
(force $X_{2}=2$) or 

\item imposing a treatment on the population for reducing testosterone levels
so that they are always low (force $X_{1}=X_{3}=1$), e.g. in an enclosed population like a prison. 
\end{enumerate}

It is easily checked that the predicted potential effect of either of these
controls under the CBN hypothesis is a plausible one. Even when the idle
system is only partially observed, the CBN hypotheses can enable us to
estimate the probable effects of such controls simply from observing a
random sample of men not subject either to a ban or a testosterone
inhibiting treatment.

The use of the CBN to express causal hypotheses has been successfully employed in many scenarios e.g. \cite{Pearl2000, Spirtesetal93}, 
while in others it is restrictive and implausible, as poignantly discussed in \cite{Shafer}.
The main problem is that causal orders are more naturally defined as refinements of a partial order on circumstances ---in a BN
represented by particular configurations of parents--- than on sets of measurements.
Again we will use the example in Section \ref{Section2violent} to demonstrate this. 
For fuller examples see \cite{Anderson, Smith07,ThwaitesPrague}.
We will omit any discussion of the important issue of exactly
how we intend to enact the control of a measurement to a particular value.    

In the example the partial order on the nodes of the BN is $X_{1},X_{2}\prec X_{3}$ and $X_{1},X_{2},X_{3}\prec X_{4}$. 
But note that, in our statement of the problem, if the man watches
the movie then by definition $X_{1}=X_{3}$.  Under this definition,
manipulating $X_{3}$ and leaving $X_{1}$ unaffected, as would be required by the CBN,
is not possible. 
If we follow the two different types of unfoldings of history: \{prior
testosterone level $X_{1}=1,2,3$, watch movie, $X_{2}=1$ posterior
testosterone level $X_{3}=1,2,3$, arrested $X_{4}=1,2$\} and \{prior
testosterone level $X_{1}=1,2,3$, don't watch movie, $X_{2}=2$, arrested $
X_{4}=1,2$\} this sort of ambiguity disappears and we could 
reasonably conjecture that these unfoldings are consistent with their
``causal order''. This might be expressed by the two context specific graphs below
\[
\begin{array}{ccccccc}
X_{1} & \rightarrow & X_{3} &\qquad & X_{1} \\ 
& \nearrow & \downarrow &\qquad & & \searrow \\ 
X_{2}=1 & \rightarrow & X_{4} &\qquad & X_{2}=2 & \rightarrow & X_{4} 
\end{array} 
\]
The joint mass function is no longer defined on the product space $S_{\boldsymbol{X}}$ with $X=\{X_1,X_2,X_3,X_4\}$.
However the joint mass function of each of these possible unfoldings is well defined and
furthermore each unfolding is expressible as a monomial in the primitive probabilities. Note that the 
class of monomials for the right-hand graph is of order one less than the left-hand one.
Many other common problems exist for which the CBN cannot express a hypothesized
causal mechanism whilst algebraic representations allows this \cite{RiccoSmith,Riccsmith07}.  

\section{Conditioning and manipulating} \label{Section3}

\subsection{Multiplication rule}
We start by fixing some notation and reviewing some known results. For a positive integer $d$ let 
$ \Delta _{d-1} = \{u\in \mathbb{R}^{d}: 
\sum_{i=1}^{d}u_{i}=1 \text{ and } u_{i}\geq 0\text{ for }i=1,\ldots ,d\}
\label{simplexnotation}
$ be 
the $(d-1)$-standard simplex 
and $ C_{d} = \{u\in \mathbb{R}^{d}:0\leq u_{i}\leq 1\text{ for }i=1,\ldots
,d\}$ the unit hypercube in $\mathbb{R}^{d}$. For a set $A\subset \mathbb{R}^{d}$, let 
$A^{\circ }$ be its interior set in the Euclidean topology.

The set of all joint probability distributions on the $n$-dimensional random vector $\boldsymbol{X}=\{X_{1},\ldots ,X_{n}\}$ taking the $r$ values in $S_{X}$,  defined in Section \ref{Section2BN},  
is identified with the $\Delta_{r-1}$ simplex
simply by listing the probabilities of each value taken by the random vector 
\[
\left( p(\boldsymbol{x}):\boldsymbol{x}=(x_{1},\ldots ,x_{n})\in S_{\boldsymbol{X}}\right)
\in \Delta _{r-1}
\]
where $p(\boldsymbol{x})=\operatorname{Prob}(\boldsymbol{X}=\boldsymbol{x})=\operatorname{
Prob}(X_{1}=x_{1},\ldots ,X_{n}=x_{n})$. 


In \cite{GarciaStillmanSturmfels2005} it is shown that independence of the random variables in $\boldsymbol{X}$ corresponds to the
requirement that $p(\boldsymbol{x})$ belongs to a Segre variety in $\Delta _{r-1}$ and that the naive Bayes model corresponds 
to the higher secant varieties of Segre varieties. While local and global independence in a BN are studied in \cite{GeigerMeekSturmfels2006}.
The most basic example, here, is that two binary random variables are independent if
$p(0,0)p(1,1)-p(1,0),p(0,1)=0$, the well known condition of zero determinant of the contingency table for $X_{1}$ and $X_{2}$.

There are various ways to map a simplex into a smaller dimensional simplex.
Some are relevant to statistics. Sturmfels (John Van Neumann Lectures 2003)
observes that, for $J\subset [n]$, marginalisation over $X_{J}$ and $X_{J^{c}}$ gives a linear map
which is a projection of convex polytopes. Namely, 
\begin{equation}
\begin{array}{lrcl}
m: & \Delta _{\boldsymbol{X}} & \longrightarrow & \Delta _{X_{J}}\times \Delta
_{X_{J^{c}}} \\ 
& (p(\boldsymbol{x}):\boldsymbol{x}\in S_{\boldsymbol{X}}) & \longmapsto & (p_{J}(x):x\in
S_{X_{j}},p_{J^{c}}(x):x\in S_{X_{j^{c}}})
\end{array}
\label{marginals}
\end{equation}
where $p_{J}(\boldsymbol{x})=\sum_{x_{i}\in \lbrack r_{i}],i\in
J^{c}}p(x_{1},\ldots ,x_{n})$ and analogously for $p_{J^{c}}(\boldsymbol{x})$. 

Here we compare the two operations of conditioning and manipulation. Diagram (\ref{Diagram}) summarises this section for binary random variables
\begin{equation}
\begin{array}{ccc}
\Delta _{2^{n}-1}^{\circ } & \longleftrightarrow & C_{2^{n}-1}^{\circ } \\ 
\downarrow &  & \downarrow \\ 
\Delta _{2^{n-1}-1}^{\circ } & \longleftrightarrow & C_{2^{n-1}-1}^{\circ } 
\end{array}
\label{Diagram}
\end{equation}

Once the order $X_{1}\prec\ldots\prec X_{n}$ is assumed on the element of a random vector $\boldsymbol{X}$ on $S_{\boldsymbol{X}}=\prod_{i=1}^n \mathbb X_i$ and $\operatorname{P}(X=\boldsymbol{x})\neq 0$ for all $\boldsymbol{x}\in S_{\boldsymbol{X}}$,  we can write
\begin{equation}
p(\boldsymbol{x})=\pi _{1}(x_{1})\pi _{2}(x_{2}|x_{1}) \ldots \pi_n(x_n|x_1,\ldots,x_{n-1})
\label{multiplicationrule}
\end{equation}
where $\pi_1(x_1)=\operatorname{Prob}(X_{1}=x_{1})$ and 
$\pi_i(x_i|x_1,\ldots,x_{i-1})=\operatorname{Prob}(X_{i}=x_{i}|X_{1}=x_{1},\ldots,X_{i-1}=x_{i-1})$ 
for $i=2,\ldots,n$.
Note that
\begin{equation*}
  \begin{array}{rcl}
    (\pi _{1}(x_{1}):x_{1}\in S_{X_{1}}) & \in  & \Delta _{r_{1}-1} \\ 
    (\pi _{2}(x_{2}|x_{1}): (x_{1},x_2) \in S_{(X_1,X_2)}  & \in  & 
    \underbrace{\Delta _{r_{2}-1}\times \ldots \times \Delta _{r_{2}-1}}_{r_1 \text{ times} } \\ 
    \vdots  &  &  \\ 
    (\pi _{n}(x_{n}|x_{1},\ldots ,x_{n-1}):(x_{1},\ldots ,x_{n})\in S_{X}) & \in 
    & \Delta _{r_{n}-1}^{\prod_{i=1}^{n-1}r_{i}}   .
  \end{array}
\end{equation*}
Hence the multiplication rule is a polynomial mapping
\begin{equation} \label{mu}
  \begin{array}{lrcl}
    \mu : & \Delta _{r_{1}-1}\times \Delta _{r_{2}-1}^{r_{2}}\dots \Delta
    _{r_{n}-1}^{\prod_{i=1}^{n-1}r_{i}} & \longrightarrow  & \Delta
    _{\prod_{i=1}^{n}r_{i}-1}
  \end{array}
\end{equation}
where the domain is parametrised by the primitive probabilities and the image space by the joint mass probabilities.
For two binary random variables let 
\begin{equation*}
  \begin{array}{rcl}
    s_{1} & = & \operatorname{Prob}(X_{1}=0) \\ 
    s_{2} & = & \operatorname{Prob}(X_{2}=0|X_{1}=0) \\ 
    s_{3} & = & \operatorname{Prob}(X_{2}=0|X_{1}=1)
  \end{array}
\end{equation*}
then $\Delta _{1}\times \Delta _{1}^{2}$ is isomorphic to $C_{3}$ and
\begin{equation*}
  \begin{array}{lrcl}
    \mu : & C_{3} & \longrightarrow  & \Delta _{3} \\ 
    & (s_{1},s_{2},s_{3}) & \longmapsto  & 
    (s_{1}s_{2},s_{1}(1-s_{2}),(1-s_{1})s_{3},(1-s_{1})(1-s_{3}))
  \end{array}
\end{equation*}
The coordinates of the image vector are listed according to a typical
order in experimental design given by taking points from top to bottom
when listed like those in Table \ref{order of points} for
$n=3$ and for binary random variables. 
\begin{table}[ht] \centering
  \begin{equation*}
    \begin{array}{ccc}
      x_{1} & x_{2} & x_{3} \\ \hline
      0 & 0 & 0 \\ 
      0 & 0 & 1 \\ 
      0 & 1 & 0 \\ 
      0 & 1 & 1 \\ 
      1 & 0 & 0 \\ 
      1 & 0 & 1 \\ 
      1 & 1 & 0 \\ 
      1 & 1 & 1
    \end{array}
  \end{equation*}
\caption{Top to bottom listings of sample points}
\label{order of points}
\end{table}

We note that the map (\ref{mu}) is \emph{not} invertible on the boundary but it
is invertible ---through the familiar equations for conditional
probability--- within the interior of the simplex where division can
be defined. For problems associated with the single unmanipulated
system this is not critical since such boundary events will occur only
with probability zero. However when manipulations are considered it
is legitimate to consider what might happen if we force the
system so that events that would not happen in the unmanipulated
system were made to happen in the manipulated system. It follows that
from the causal modelling point of view the conditional
parametrisation is more desirable.

\subsection{Conditioning as a projection}
Consider $i\in [ n]$ and define $x_{-i}=(x_{1},\ldots,x_{i-1},x_{i+1},\ldots ,x_{n})$ and  
$[r_{-i}]=\mathbb X_1 \times \ldots \times \mathbb X_{i-1} \times \mathbb X_{i+1} \times\ldots \times \mathbb X_n$. 
Analogous symbols are defined for $J\subset [n]$.
For 
  $x_{i}^{\ast }\in [r_{i}]$ such that $\operatorname{Prob}(X_{i}=x_{i}^{\ast })\neq
0$,  the conditional probability of $\boldsymbol{X}$ on
$\{X_{i}=x_{i}^{\ast }\}$ is defined as
\begin{equation*}
  \operatorname{Prob}(\boldsymbol{X}=\boldsymbol{x}|X_{i}=x_{i}^{\ast})
  =\left\{ 
    \begin{array}{ll}
      0 & \text{if }x_{i}\neq x_{i}^{\ast } \\ 
      \displaystyle\frac{p(\boldsymbol{x})}{\sum_{x_{-i}\in 
          [r_{-i}]}p(\boldsymbol{x})} & \text{if }x_{i}=x_{i}^{\ast}  .
    \end{array}
  \right. 
\end{equation*}
Outside the set $ x_{i}\neq x_{i}^{\ast}$, this mapping is an example of
the simplicial projection on the face $x_{i}=x_{i}^{\ast}$.
Briefly, any simplex $ \Delta $ in the Euclidean space is the join of
any two complementary faces, which are  simplices themselves.
In particular, if $F$ and $F^{c}$ are complementary faces, then
each point $P$ in the simplex and not in $F$ or $ F^{c}$ lies on the
segment joining some point $P_{F}$ in $F$ and some point $ P_{F^{c}}$
in $F^{c}$, and on only one such segment. This allows us to define a
projection $\pi _{F}:\Delta \setminus F^{c}\rightarrow F$, by $\pi
_{F}(P)=P_{F}$ if $P\notin F$ and $\pi _{F}(P)=P$ if $P\in F$.
\begin{example}  \normalfont{For $n=2$ and $P=(p(0,0),p(0,1),p(1,0),p(1,1))$ with
    $p(0,0)+p(0,1)\neq 0$, $F=\{x\in \Delta_3: x=(x_1,x_2,0,0) \}$ and
    $F^c=\{x\in \Delta_3: x=(0,0,x_3,x_4) \}$, we have
    \begin{equation*} \begin{array}{rcl}
        P_F &=& \displaystyle \frac1{p(0,0)+p(0,1)}(p(0,0),p(0,1), 0,0) \\
        P_{F^c} &=& \displaystyle \frac1{p(1,0)+p(1,1)}(0,0,p(1,0),p(1,1)) \\
        P &=& (p(0,0)+p(0,1)) P_F+ (p(1,0)+p(1,1)) P_{F^c}  .  
      \end{array} \end{equation*}
    For $X$ and $Y$ binary random variables, the operation $P(Y|X=0)$ corresponds to  
    \begin{equation*}
      \begin{array}{rcl} \Delta_{3}^{\circ} & \longrightarrow &
        \Delta_{1}^{\circ}  \\
        (p(0,0),p(0,1),p(1,0),p(1,1)) & \longmapsto &
        \displaystyle \frac1{p(0,0)+p(0,1)}(p(0,0),p(0,1))
      \end{array}
    \end{equation*}
    It can be extended to the boundary $ \Delta_{1} $ giving for
    example the probabilities mass functions for which $p(0,0)=0$ or
    $1$.  } \end{example}

By repeated projections we can condition on
$\operatorname{Prob}(X_{J}=x_{J}^{\ast})>0$ with
$J\subset \lbrack n]$.     
Then, the operation of conditioning returns a ratio of
polynomial forms of the type $x/(x+y+z)$ where $x,y,z$ stand for joint
mass function values. This has been implemented in computer algebra softwares by various researchers, as an application of elimination theory. 
A basic algorithm considers  indeterminates $t_{\boldsymbol{x}}$ with 
$\boldsymbol{x}\in S_{\boldsymbol{X}}$ for the domain space and $b_{\boldsymbol{y}}$ with $\boldsymbol{y}\in [r_{-J}]$ 
for the image space. 
The joint probability
mass $(p(\boldsymbol{x}):\boldsymbol{x}\in S_{\boldsymbol{X}})$ corresponds to $I=\operatorname{Ideal}(t_{\boldsymbol{x}}-p(\boldsymbol{x}):\boldsymbol{x}\in
S_{\boldsymbol{X}})$ of $\mathbb{Q}[t_{\boldsymbol{x}}:\boldsymbol{x}\in S_{\boldsymbol{X}}]$, 
the set of polynomials in the $t_{\boldsymbol{x}}$ with rational coefficients.
 Its projection onto the face $F_{J}$ can be computed by elimination as follows by adjoining a dummy
indeterminate $l$ and viewing $I$ as an ideal in $\mathbb{R}[t_{\boldsymbol{x}}:\boldsymbol{x}\in
S_{\boldsymbol{X}},b_{\boldsymbol{y}}:\boldsymbol{y}\in \lbrack r_{-J}],l]$. Consider $I+J$ where $J$ is the ideal
generated by 
\begin{equation}
\begin{array}{l}
l-\sum_{\boldsymbol{y}\in \lbrack r_{-J}]}b_{\boldsymbol{y}} \\ 
b_{\boldsymbol{y}}l-p(\boldsymbol{x})\sum_{\boldsymbol{y}\in \lbrack r_{-J}]}b_{\boldsymbol{y}}
\end{array}
\label{sumideals}
\end{equation}
where $\boldsymbol{x}$ and $\boldsymbol{y}$ are suitably matched by the definition of conditioning. 
Then the elimination ideal of $I+J$
of the $l$ and $t_{\boldsymbol{x}}$ variables corresponds to the simplicial projection.

\begin{example}
\normalfont{ We use the freely available software \cocoa \cite{CoCoA} to project the point  
    $P=(1/3,1/3,1/3)\in \Delta_2$ onto the face $x_1+x_2=1$. The ideal of the point $P$ in the
    $t[1],t[2],t[3]$ indeterminates is  $I=\operatorname{Ideal}(t[1]-1/3, t[2]-1/3, t[3]-1/3)$. 
    $J$ describes a plane parallel to the face $x[3]=0$ of
      the simplex and $J$ is the ideal in Equation
      (\ref{sumideals}). \texttt{Lex} and \texttt{GBasis} are the
      technical commands to perform the elimination. The result is in the last line.
      \begin{small} \begin{verbatim}
 Use T::=Q[t[1..3]ls[1..2]],Lex;
 I:=Ideal(t[1]-1/3,t[2]-1/3,t[3]-1/3);
 L:=t[1]+t[2]-l;
 J:=Ideal(s[1] l-1/3, s[2] l-1/3, s[1]+s[2]-1,L, s[1]+s[2]-1);
 GBasis(I+J);
 [t[3] - 1/3, t[2] - 1/3, t[1] - 1/3, 
  s[1] + s[2] - 1, -l + 2/3, 2/3s[2] - 1/3]
 \end{verbatim} \end{small}
} \end{example}

\subsection{The manipulation of a Bayesian network}\label{Section3.3}

In Equation (3.10) of \cite{Pearl2000} J. Pearl, starting from a joint probability mass function
on $\boldsymbol{X}$, an $x_{i}^{\ast}$ value and assuming a causal order for a BN, defines a new probability mass
function for the intervention $X_{i}=x_{i}^{\ast}$. 
In general, we 
partition $[n]=\{i\}\cup \{1,\ldots ,i-1\}\cup \{i+1,\ldots ,n\}$
and assume this partition compatible with a causal order on $\boldsymbol{X}$, that is if 
$j\in \{1,\ldots ,i-1\}$ then $X_{j}$ is not affected by the intervention on $X_{i}$. 
If the probabilistic structure on $\boldsymbol{X}$ is a BN then we consider a regular BN. 
We consider the parametrization 
\[
p(\boldsymbol{x})=p(x_{1},\ldots ,x_{i-1})p(x_{i}|x_{1},\ldots
,x_{i-1})p(x_{i+1},\ldots ,x_{n}|x_{1},\ldots ,x_{i})
\]
for which a probability is seen as a point in 
\[
\Delta _{\lbrack r_{i-1}]-1}\times \Delta
_{r_{i}-1}^{\prod_{j=1}^{i-1}r_{j}}\times \Delta
_{\prod_{j=i+1}^{n}r_{j}-1}^{\prod_{j=1}^{i}r_{j}} .
\]
The intervention or manipulation operation is defined only for image points for which $x_{i}\neq
x_{i}^{\ast}$ and returns a point in 
\[
\Delta _{\lbrack r_{i-1}]-1}\times \Delta
_{\prod_{j=i+1}^{n}r_{j}-1}^{\prod_{j=1}^{i-1}r_{j}}
\]
namely the point with coordinates 
\[
p(x_{1},\ldots ,x_{i-1})\text{ and }p(x_{i+1},\ldots ,x_{n}|x_{1},\ldots
,x_{i}^{\ast })  
\]
for $(x_{1},\ldots ,x_{i-1})\in \mathbb X_1\times\ldots \times \mathbb X_{i-1}$ and 
$(x_{i+1},\ldots ,x_{n}) \in \mathbb X_{i+1}\times\ldots \times \mathbb X_{n}$.
Note that this map is naturally defined over the boundary. 
In contrast there is no unique map extendible to the boundary of the probability space in $\Delta_{\boldsymbol{X}}$. 

For binary random variables it is the orthogonal projection from $C_{2^{n}-1}$ 
onto the face $x_{i}\neq x_{i}^{\ast}$ which is
identified with the hypercube $C_{2^{n-1}-1}$.
In general, for a regular BN this is an orthogonal projection in the associated conditional parametrisation, 
which then seems the best parametrization in which to perform computations. 
The post manipulation joint mass function on $\boldsymbol{X}\setminus X_i$ is then 
$
p(x_{1},\ldots ,x_{i-1}) p(x_{i+1},\ldots ,x_{n}|x_{1},\ldots
,x_{i}^{\ast })  
$ which, factorised in primitive probabilities, gives a monomial of degree $n-1$, one less than in Equation (\ref{BNfact}).
In this sense, under the conditional parametrization, the effect of a manipulation or control gives a much simpler
algebraic map than the effect of conditioning.

Its formal definition depends only on the causal order, the second bullet point in Section \ref{Section2BN}, 
and not on the probabilistic structured of the BN. In particular it does not depend on the homogeneity of the factorization of the joint mass function on $\boldsymbol{X}$ across all settings.  
This observation allowed us to extend this notion to larger classes of discrete causal models. See 
\cite{RiccoSmith,Riccsmith07} and Section \ref{Section4}. 

Identification problems associated with the estimation of some probabilities after manipulation from passive observations (manifest variables measured in the idle system) have been formulated as an elimination problem in computational commutative algebra. For example in the case of BN the case study in \cite{kuroki2007},
giving a graphical application of the back-door theorem \cite{Pearl2000}, has been replicated algebraically by Matthias Drton using the parametrization in primitive probabilities. Ignacio Ojeda addresses from an algebraic view point a different and more unusual identification problem in a causal BN with four nodes. He uses the $p(\boldsymbol{x})$ parameters and the description of the BN as a toric ideal. Both are personal communications at the workshop to which this volume is dedicated.

In general, a systematic implementation of these problems in computer algebra softwares will be slow to run.  
At times some pre-processing can be performed in order to exploit the symmetries and
invariances to various group action for certain classes of statistical
models \cite{Mond}. Other times a re-parametrisation in terms of
non-central moments loses an order of magnitude effect on the speed of
computation  \cite{SettimiSmith2000} and hence can be useful. 
Nevertheless in this algebraic framework  many non-graphically based symmetries which appear in common
models are much easier to exploit than in a graphical setting. This suggests that the algebraic
representation of causality is a promising way of computing the
identifiability of a causal effect in much wider classes of models than BN. 

\section{Reformulating causality algebraically} \label{Section4} 
To recap: 
\begin{enumerate}
\item \label{order} a total order on $\boldsymbol{X}=\{X_1,\ldots,X_n\}$ and an associated multiplication rule as in Equation (\ref{multiplicationrule}) are fundamental. These determine a set of primitive probabilities;
\item \label{studeny} a discrete BN can be described through a set of linear equations equating primitive probabilities, Equations (\ref{BNeqs}), together with inequalities to express non negativity of probabilities and linear equations for the sum-to-one constraints; 
\item a BN is based on the assumption that the factorization in Equation (\ref{BNfact}) holds across all values of $\boldsymbol{x}$ in a cross product sample space. Recall that in \cite{SettimiSmith2000} it is shown that identification depends on the sample space structure, in particular on the number of levels a variable takes; 
\item within a graphical framework subsets of whole variables in $\boldsymbol{X}$ are considered manifest or hidden;
\item \label{causalcontrol} mainly the causal controls being studied in e.g. \cite{Pearl2000,Spirtesetal93} correspond to setting subsets of variables in $\boldsymbol{X}$ to take particular values 
and often the effect of a cause is expressed as a polynomial function of the primitive probabilities, in particular the probability of a suitable marginal;
\item \label{identpoint} identification problems formulated in the graphical framework of a BN and intended as the writing of an effect of a cause in terms of manifest variables are basically elimination problems. Hence they can be addressed using elimination theory from computational commutative algebra. In particular 
theorems like the front-door theorem and the back-door theorem are proved using clever algebraic eliminations, see \cite{Pearl2000}. 
 \end{enumerate}
 
The above scheme can be modified in many directions to include non-graphical models and causal functions not expressible in a graphical framework, like those in Section \ref{SectionCausal}. Identification problems can still be addressed with algebraic methods as in Item \ref{identpoint} above. An indispensable point for a causal interpretation of a model is a partial order either on $\boldsymbol{X}$ or on $S_{\boldsymbol{X}}$, where the sample space may be generalised to be not of product form. 

A first generalisation is in \cite{Riccimpu} where the authors substitute the binomials in Item~\ref{studeny} above with linear equations and the inequalities in Item~\ref{order} with inequalities between linear functions in the primitive probabilities. If there exists at least a probability distribution over $\boldsymbol{X}$ satisfying this set of equations and inequalities then the model is called a feasible Bayesian linear constraint model. 

Of course a mere algebraic representation of a model will lose 
the expressiveness and interpretability associated with the compact topology of most graphical structures and hence to dispense completely with the graphical constraints might not always be advisable. But a combined use of a graphical representation and an algebraic one will certainly allow the formulation of more general model classes and will allow causality to benefit of computational and interpretative techniques of algebraic geometry as currently happens in computational biology \cite{PachterSturmfels2005}. A causal model structure based on a single rooted tree and amenable of an algebraic formulation is studied in \cite{RiccoSmith,Riccsmith07}.
In there, following \cite{Shafer} the focus of the causal model is shifted from the factors in $\boldsymbol{X}$ to the actual circumstances. Each node of the tree represents a ``situation'' ---in the case of a BN a possible setting of the $\boldsymbol{X}$ vector--- and the partial order intrinsic to the tree is consistent with the order in which we believe things can happen. This approach has many
advantages, freeing us from the sorts of ambiguity discussed in Section \ref{Section2violent} and allowing us to define simple causal controls that enact a particular policy \emph{only} when conditions might require that control. 
 
\subsection{Causality based on trees} \label{SectionTree}
Assume a single rooted tree $\mathcal T=(V, E)$ with vertex set $V$ and edge set $E$. Let $e=(v,v^\prime)$ be a generic edge from $v$ to $v^\prime$ and associate to $e$ a possibly unknown transition probabilities $\pi(v^\prime|v)\in [0,1]$ under the constraint $\sum_{v^\prime: (v,v^\prime)\in E} \pi(v^\prime|v)=1$, for all $v\in V$ which are not leaf vertices. The set $\Pi=\{\pi(v^\prime|v)\}$ gives a parametrization of our model and the $\pi(v^\prime|v)$ are called primitive probabilities. Let $\mathbb X$ be the set of root-to-leaf paths in $\mathcal T$ and for $\lambda=(e_1,\ldots,e_{n(\lambda)})=(v_0,\ldots,v_{n(\lambda)})\in \mathbb X$, where $v_0$ is the root vertex and $v_{n(\lambda)}$ a leaf vertex, define the polynomials 
\begin{equation} \label{treefact}
{p}(\lambda)=\prod_{i=0}^{n(\lambda)-1} \pi(v_{i+1}|v_i) .
\end{equation}
In \cite{RiccoSmith} it is shown that $(\mathbb X, 2^\mathbb X, {p}(\cdot))$ is a probability space. The set of circumstances of interest is then represented by the nodes of the tree and the probabilistic events are given by the leaves of the tree, equivalently the root-to-leaf paths. 

Here are three examples from the literature. Once an order on $\boldsymbol{X}$ has been chosen, a BN corresponds to a tree whose root-to-leaf paths have all the same length, $S_{\boldsymbol{X}}=\mathbb X$ and its independence structure is translated into equalities of some primitive probabilities \cite{Smith07,Shafer}. 
The basic saturated model individuated by the polynomials in Equations (\ref{treefact}) augmented with a set of algebraic equations in the elements of $\Pi$ has been called algebraically constraint tree in \cite{Riccsmith07}. 
In \cite{RiccoSmith,Smith07,ThwiatesParis,ThwaitesPrague} a model based on a tree and called a chain event graph has now been developed and explored to some level of detail. 

There is a natural partial order associated with the tree which can be used as a framework to express causality: $v\prec v^{\prime }$ if there exists $\lambda\in \mathbb X$ such that $v,v^{\prime }\in \lambda$ and $v$ lies closer to $v_0$ than $v^{\prime }$. A tree is \emph{regular}
if in the problem we are modelling the circumstance represented by $v$ occurs before the one represented by $v^{\prime }$ whenever $v\prec v^{\prime }$. The effects of a control on a regular tree $T$ can now be defined in total analogy to Item \ref{causalcontrol} above by modifying the values of some primitive probabilities or more generally by defining constraints in the primitive probabilities that have a causal interpretation. 

\begin{definition} \label{manipulation}
Let $\mathcal{T}=(V,E)$ be a regular tree and $\Pi$ the associated primitive probabilities.
A {\normalfont manipulation} of the tree is given by a subset $F\subset E$ and 
an extra set of parameters associated to edges in $F$, 
namely  
$\widehat{\Pi}_F=\{ \widehat{\pi}(v^\prime|v): (v,v^\prime)\in F \}$
under the constraints 
 $\widehat{\pi}(v^\prime|v)\geq 0$  for all $(v,v^\prime)\in F$ 
and $\sum_{v^\prime: (v,v^\prime)\in E\setminus F }{\pi}(v^\prime|v)+\sum_{v^\prime: (v,v^\prime)\in F }\widehat{\pi}(v^\prime|v)=1$. 
Furthermore the $\widehat{\pi}(v^\prime|v)$ are assumed to be functions of the primitive probabilities for all $(v,v^\prime)\in F$. 
\end{definition}

For example in the typical manipulations in \cite{Pearl2000,RiccoSmith} and in Section \ref{Section3.3} some $\widehat{\pi}(v^\prime|v)$ are chosen equal to one and hence some others equal to zero. Here we observe that Definition \ref{manipulation} translates into a  map similar to the one discussed for BN's in Section \ref{Section3.3}. 

To simplify notation let $S\subset V$ be the set of non leaf vertices in $\mathcal T$. For $v\in S$ let $\mathbb X(v)=\{ v^\prime\in V : (v,v^\prime) \in E \}$ and $r_v$ be the cardinality of $\mathbb X(v)$. Then the saturated model on a tree is equivalent to the list of primitive probabilities $\boldsymbol{\pi}=(\pi(v^\prime|v): (v,v^\prime)\in E) \in \prod_{v\in S} \Delta_{r_v-1}$ together with the semi-algebraic constraints $\sum_{v^\prime: (v,v^\prime)\in E} \pi(v^\prime|v)=1$ and $\pi(v^\prime|v)\geq 0$ and with the partial order of the tree, equivalently Equations (\ref{treefact}).

For $F\subset E$ let $D_F=\{v\in V: \text{there exists } v^\prime \text{ such that }(v,v^\prime)\in F\}$.  
We can re-arrange the list $\boldsymbol\pi$ to list first primitive probabilities of edges not in $F$ and then a manipulation on $F$ is given by the mapping
\begin{multline*}
\prod_{v\in S\setminus D_F} \Delta_{r_v-1} \times \prod_{v\in D_F} \Delta_{r_v-1} 
  \longrightarrow
  \prod_{v\in S\setminus D_F} \Delta_{r_v-1} \times \prod_{v\in D_F} \Delta_{r_v-1}
  \\
  (\pi(v^\prime|v): (v,v^\prime)\in E) 
 \longmapsto 
 (\pi(v^\prime|v): (v,v^\prime)\in E\setminus F, \widehat\pi(v^\prime|v): (v,v^\prime)\in F ) 
\end{multline*}
For the typical manipulations in \cite{Pearl2000,RiccoSmith} and in Section \ref{Section3.3} 
this map simplifies to an orthogonal projection on $\prod_{v\in S\setminus D_F} \Delta_{r_v-1}\ni (\pi(v^\prime|v): v\in S\setminus D)$.  

\subsection{Extreme causality} \label{extreme}

To effectively discuss causal maps we notice that we need 1. a finite set of ``circumstances'' ---in the BN represented by parent configurations and in the tree by the tree situations--- augmented with a finite set of ``terminal circumstances'', e.g. the possible final outcomes of an experiment, and 2. a partial order defined on these circumstances expressing the causal hypotheses of the system.
The circumstances could be identified with particular types of causally critical events in the event space of the uncontrolled system, e.g. $\mathbb X$ of Section \ref{SectionTree}. 

Hence let $V=\{v\}$ be the finite set representing circumstances and terminal circumstances and  $\prec$ a partial order on $V$. The partial order can be visualised through its Hasse diagram and corresponds to a finite number of chains of elements of $V$. A chain is a list of elements in $V$: $\lambda=(v_1,\ldots,v_n)$ where  $v_{i-1}\prec v_{i}$ for all $i=2,\ldots,n$ and such that for no $v^\prime,v^{\prime\prime}\in V$ we have $v^\prime\prec v_1$ and $v_n\prec v^{\prime\prime}$. 
A circumstance can belong to more than one chain and chains can have different lengths, initial circumstances and terminal circumstances. A chain represents a possible unfolding of the problem we are modelling, from a starting point, $v_0$, to an end point, $v_n$. The order represents the way circumstances succeed one another and one could be the cause of a subsequent one. 

Once the partial order in $V$ has been elicited, a parametrization of a saturated statistical model on $V$ 
can be defined as a set of transition probabilities:
$\pi(v^\prime|v)\in [0,1]$ where $v,v^\prime\in V$ are such that $v^\prime$ and $v$ are in the same chain, say $\lambda$,
$v\prec v^\prime$ and there is no $v^\ast\in \lambda$ such that $v\prec v^\ast \prec v^\prime$. That is, there is a chain to which both $v$ and $v^\prime$ belong and $v$ precedes $v^\prime$ immediately in the chain. 
We call $\pi(v^\prime|v)$ primitive probabilities, collect them in a vector $\boldsymbol \pi=(\pi(v^\prime|v))$ and note that they can be given as labels to the edges of the Hasse diagram.
Moreover, we require that if $v$ belongs to more than one chain, then the sum of the transition probabilities $\pi(\cdot|v)$ is equal to one, i.e. $\sum_{v^\prime\in \lambda: v\in \lambda} \pi(v^\prime|v)=1$. This defines the domain space of $\boldsymbol \pi$ as a product of the simplices in total analogy to the cases of BN's and trees.  
The probability of a chain $\lambda$ is now defined as $p(\lambda)=\prod_{i=1}^{n} \pi(v_{i}|v_{i-1})$, in analogy to Equations (\ref{treefact}) and (\ref{multiplicationrule}). 

Thus, we have determined a saturated model parametrised with $\boldsymbol\pi$ and given by the sum-to-one constraints and the non-negative conditions.
A sub-model, say $S$, can be defined by adjoining equalities and inequalities between polynomials or ratios of polynomials in the primitive probabilities, say $q(\boldsymbol\pi)=0$ and $r(\boldsymbol\pi)>0$, where $q$ and $r$ are polynomials or ratios of polynomials. Of course one must ensure that there is at least one solution to the obtained system of equalities and inequalities; that is, that the model is feasible. Sub-models can also be defined through a refinement of the partial order.

Next, causality can be defined implicitly by considering a set $F$ of edges of the Hasse diagram and for $(v,v^\prime)\in F$ adjoining to $S$ a new set of primitive probabilities $\widehat{\pi}(v^\prime|v)$ and some equations 
$\widehat{ \pi}(v^\prime|v) = f_{(v,v^\prime)}(\boldsymbol\pi)$ where $f_{(v,v^\prime)}$ is a polynomial. Collect the new parameters in the list $\widehat{\boldsymbol{\pi}}=(\widehat{\boldsymbol{\pi}}(v^\prime|v))=f(\boldsymbol\pi)$, where $f=(f_{(v,v^\prime)}: (v,v^\prime)\in F)$.

Identifiability problems are now formulated as in previous sections. Suppose we observe some polynomial equalities of the primitive probabilities, $m=m(\boldsymbol{\pi})$, and even some inequalities $m(\boldsymbol{\pi})>0$, where $m$ is a vector of polynomials. 
Then we are interested in checking whether a total cause, $e=e(\widehat{\boldsymbol{\pi}})$, is identifiable from and compatible with the given observation. This computation could be done by using techniques of algebraic geometry in total analogy to BN's and trees as discussed in Item~\ref{identpoint}. 

The top-down scheme in Table \ref{summary} summarises all this. 
In the top cell we have a semi-algebraic set-up involving equalities and inequalities in the $\boldsymbol{\pi}$ parameters involving polynomials or ratios of polynomials.
We must ensure that the set of values of $\boldsymbol{\pi}$ which solve this system of equalities and inequalities is not empty, i.e. the model is feasible. 
In the next two cells we add two sets of indeterminates: $\widehat{\boldsymbol{\pi}}$ and $\boldsymbol{m}=(m)$, and some equalities and inequalities of polynomials in the $\boldsymbol\pi$. Then the effect $e$ is uniquely identified if there is a value $\boldsymbol\pi^\ast$ of $\boldsymbol\pi$ satisfying the system and $e=e(\boldsymbol{m}(\boldsymbol\pi^\ast))$. 

\begin{table}[t]\label{summary}
\begin{tabular}{|ll|}
\hline
{Saturated model} & $0\leq \pi(v^\prime|v) \leq 1$ and $\sum_{v^\prime\in \lambda: v\in \lambda} \pi(v^\prime|v)=1$ \\
{Submodel} & $q(\boldsymbol\pi)=0$ and $r(\boldsymbol\pi)>0$ \\ 
\hline
{System manipulation} & $\boldsymbol{\widehat{ \pi} }= f(\boldsymbol\pi)$  \\
\hline
{Manifest} & $m=m(\boldsymbol{\pi})$ and $n(\boldsymbol{\pi})>0$ \\
\hline
{Identifiability} & $e=e(\boldsymbol{m}(\boldsymbol\pi^\ast))$\\
\hline
\end{tabular}
\caption{Summary of Section \ref{extreme}}
\end{table}

All the models considered in this paper fall within this framework and within the class of algebraic statistical models \cite{DrtonSullivant2007}. 
In particular in CEG models \cite{Smith07} circumstances are defined as sets of vertices of a
tree and the partial order is inherited from the tree order. CEG's in a causal context have been studied in \cite{Riccsmith07}
and they have been  applied to the study of biological
regulation models \cite{Anderson}. 
We conjecture that there are many other classes of causal models that have an algebraic formulation of this type and 
are useful in practical applications. 
We end
this paper by a short discussion of how the identifiablity issues associated
with the non-graphical example of Section \ref{Section2violent} can be addressed
algebraically.

\subsection{Identifying a cause in our example} \label{Section4.1} 
For the example in Section \ref{Section2violent} assume 
conditions (\ref{nomovie}) and (\ref{movietable}). Hence, for $x_{1}=1,2,3$ the non-zero probabilities associated
with not viewing the movie are
$p(x_{1},2,x_{1},1) = \pi _{1}(x_{1})\pi _{2}(2)\pi _{4}(1|2,x_{1})$ and 
$p(x_{1},2,x_{1},2) = \pi _{1}(x_{1})\pi _{2}(2)\pi _{4}(2|2,x_{1})$
whilst the probabilities associated with viewing it are given in Table \ref{malaimpaginazione}.
\begin{table} \label{malaimpaginazione} \begin{eqnarray*}
p(1,1,1,1) &=&\pi _{1}(1)\pi _{2}(1)\pi _{3}(1|1,1)\pi _{4}(1|1,1) \\
p(1,1,1,2) &=&\pi _{1}(1)\pi _{2}(1)\pi _{3}(1|1,1)\pi _{4}(2|1,1) \\
p(1,1,2,1) &=&\pi _{1}(1)\pi _{2}(1)\pi _{3}(2|1,1)\pi _{4}(1|1,2) \\
p(1,1,2,2) &=&\pi _{1}(1)\pi _{2}(1)\pi _{3}(2|1,1)\pi _{4}(2|1,2) \\
p(1,1,3,1) &=&\pi _{1}(1)\pi _{2}(1)\pi _{3}(3|1,1)\pi _{4}(1|1,3) \\
p(1,1,3,2) &=&\pi _{1}(1)\pi _{2}(1)\pi _{3}(3|1,1)\pi _{4}(2|1,3) \\
p(2,1,2,1) &=&\pi _{1}(2)\pi _{2}(1)\pi _{3}(2|1,1)\pi _{4}(1|1,2) \\
p(2,1,2,2) &=&\pi _{1}(2)\pi _{2}(1)\pi _{3}(2|1,1)\pi _{4}(2|1,2) \\
p(2,1,3,1) &=&\pi _{1}(2)\pi _{2}(1)\pi _{3}(3|1,1)\pi _{4}(1|1,3) \\
p(2,1,3,2) &=&\pi _{1}(2)\pi _{2}(1)\pi _{3}(3|1,1)\pi _{4}(2|1,3) \\
p(3,1,3,1) &=&\pi _{1}(3)\pi _{2}(1)\pi _{3}(3|1,1)\pi _{4}(1|1,3) \\
p(3,1,3,2) &=&\pi _{1}(3)\pi _{2}(1)\pi _{3}(3|1,1)\pi _{4}(2|1,3)  .
\end{eqnarray*}
\caption{Probabilities associated with viewing the movie} \end{table}

Consider the two controls described in the bullets in Section \ref{SectionCausal}. The first, banning the film, gives non-zero probabilities 
for $x_{1}=1,2,3$ satisfying the equations
$\widehat{p}(x_{1},2,x_{1},1) = \pi _{1}(x_{1})\pi _{4}(1|2,x_{1}) $ and
$\widehat{p}(x_{1},2,x_{1},2) = \pi _{1}(x_{1}))\pi _{4}(2|2,x_{1})$.
The second,  the fixing of testosterone levels to low for all time, gives
manipulated probabilities 
\begin{eqnarray*}
&\widehat{\widehat{p}}(1,2,1,1) =\pi _{2}(2)\pi _{4}(1|2,1) \qquad
&\widehat{\widehat{p}}(1,2,1,2) =\pi _{2}(2)\pi _{4}(2|2,1) \\
&\widehat{\widehat{p}}(1,1,1,1) =\pi _{2}(1)\pi _{4}(1|1,1) \qquad
&\widehat{\widehat{p}}(1,1,1,2) =\pi _{2}(1)\pi _{4}(2|1,1)  .
\end{eqnarray*}
Now consider three experiments. Experiment~1 of Section \ref{SectionManifest} exposes men to the movie,  measuring their
testosterone levels before and after viewing the film. This obviously
provides us with estimates of $\pi _{1}(x_{1})$, for $x_{1}=1,2,3$ and $\pi
_{3}(x_{3}|1,x_{1})$ $1\leq x_{1}\leq x_{3}\leq 3$. Under Experiment~2 of Section \ref{SectionManifest} a
large random large sample is taken over the relevant population providing
good estimates of the probability of the margin of each pair of $X_{2}$ 
and the level of testosterone $X_{3}$ on those who fought, $\{X_{4}=1\}$, but
only the probability of not fighting otherwise. 
So you can estimate the values of and sample for $x_{1}=1,2,3$
$
p(x_{1},2,x_{1},1)=\pi _{1}(x_{1})\pi _{2}(2)\pi _{4}(1|2,x_{1})
$
and
\begin{eqnarray*}
p(1,1,1,1) &=&\pi _{1}(1)\pi _{2}(1)\pi _{3}(1|1,1)\pi _{4}(1|1,1) \\
p(1,1,2,1) &=&\pi _{1}(1)\pi _{2}(1)\pi _{3}(2|1,1)\pi _{4}(1|1,2) \\
p(1,1,3,1) &=&\pi _{1}(1)\pi _{2}(1)\pi _{3}(3|1,1)\pi _{4}(1|1,3) \\
p(2,1,2,1) &=&\pi _{1}(2)\pi _{2}(1)\pi _{3}(2|2,1)\pi _{4}(1|1,2) \\
p(2,1,3,1) &=&\pi _{1}(2)\pi _{2}(1)\pi _{3}(3|2,1)\pi _{4}(1|1,3) \\
p(3,1,3,1) &=&\pi _{1}(3)\pi _{2}(1)\pi _{3}(3|3,1)\pi _{4}(1|1,3)  .
\end{eqnarray*}%
Note the last probability is redundant since it is one minus the sum of
those given above. Finally  Experiment 3 is a survey that informs us about
the proportion of people watching the movie on any night, i.e tells us 
$\left( \pi _{2}(1),\pi _{2}(2)\right) $.

Now suppose we are interested in the total cause \cite{Pearl2000}
\[
e=\sum_{x_{1},x_{3}}\widehat{p}(x_{1},2,x_{3},1)=\sum_{x_{1}}\pi _{1}(x_{1})\pi _{4}(1|2,x_{1})
\]
of fighting if forced not to watch. Clearly this is identified from an 
experiment that includes Experiments~2 and~3 by summing and division by $\pi
_{2}(2)$, but by no other combination of experiments. 
Similarly $e^\prime=\widehat{\widehat{p}} (1,1,1,1)=\pi _{2}(1)\pi _{4}(1|1,1)$, the probability a man with
testosterone levels held low watches the movie and fights, is identified from
$p(1,1,1,1)$ obtained from Experiment~1 and~2 by division. 

The movie example falls within the general scheme of Section \ref{Section4}. 
Of course a graphical representation of the movie example, e.g. over a tree or even a BN, is possible and
useful. But one of the point of this paper is to show that 
when discussing causal modelling the first step does not need to be the elicitation of a graphical structure whose geometry can then be examined through its underlying algebra. Rather an algebraic formulation based on the identification of the circumstances of interest, e.g. the set $V$, and the elicitation of a causal order, e.g. the partial order on $V$, is a more naturally starting point. Clearly in such framework on one hand the graphical type of symmetries embedded and easily visualised on e.g. a BN are not immediately available but they can be retrieved (for an example involving CEG and BN see \cite{Smith07}). On the other hand algebraic type of symmetries might be easily spotted and be exploited in the relevant computations.

In this example computation was simple algebraic operation while in more complex case we might need to recur to a computer.
Of course the usual difficulties of using current computer code for elimination problems of this kind remain, 
because inequality constraints are not currently integrated into
software and because of the high number of primitive probabilities involved. 
Caveats in Section \ref{Section3} for BN's, like the advantages of ad-hoc parametrizations, apply to these structures based on trees and/or defined algebraically. 
 
\section{Acknowledgements}
This work benefits from many discussions with various colleagues. In particular we acknowledge gratefully Professor David Mond for helpful discussion on the material in Section \ref{Section3} and an anonymous referee of a related paper for a version of our main example.

\end{document}